\documentclass[twocolumn]{aastex61}

\usepackage{natbib}
\usepackage{amsmath}

\def\lsim{\hbox{\rlap{\raise 0.425ex\hbox{$<$}}\lower 0.65ex\hbox{$\sim$}}}
\def\gsim{\hbox{\rlap{\raise 0.425ex\hbox{$>$}}\lower 0.65ex\hbox{$\sim$}}}

\def\arcdeg{\mbox{$^\circ$}}

\shorttitle{Unprecedented Properties of SSS17a}
\shortauthors{Siebert et~al.}

\begin{document}

\title{The Unprecedented Properties of the First Electromagnetic Counterpart to a Gravitational Wave Source}

\author{M.~R.~Siebert}
\affiliation{Department of Astronomy and Astrophysics, University of California, Santa Cruz, CA 95064, USA}

\author{R.~J.~Foley}
\affiliation{Department of Astronomy and Astrophysics, University of California, Santa Cruz, CA 95064, USA}

\author{M.~R.~Drout}
\affiliation{The Observatories of the Carnegie Institution for Science, 813 Santa Barbara Street, Pasadena, CA 91101}
\affiliation{Hubble and Carnegie-Dunlap Fellow}

\author{C.~D.~Kilpatrick}
\affiliation{Department of Astronomy and Astrophysics, University of California, Santa Cruz, CA 95064, USA}

\author{B.~J.~Shappee}
\affiliation{The Observatories of the Carnegie Institution for Science, 813 Santa Barbara Street, Pasadena, CA 91101}
\affiliation{Hubble and Carnegie-Princeton Fellow}
\affiliation{Institute for Astronomy, University of Hawai\'{i}, 2680 Woodlawn Drive, Honolulu, HI 96822, USA}

\author{D.~A.~Coulter}
\affiliation{Department of Astronomy and Astrophysics, University of California, Santa Cruz, CA 95064, USA}

\author{D.~Kasen}
\affiliation{Nuclear Science Division, Lawrence Berkeley National Laboratory, Berkeley, CA 94720, USA}
\affiliation{Departments of Physics and Astronomy, University of California, Berkeley, CA 94720, USA}

\author{B.~F.~Madore}
\affiliation{The Observatories of the Carnegie Institution for Science, 813 Santa Barbara Street, Pasadena, CA 91101}

\author{A.~Murguia-Berthier}
\affiliation{Department of Astronomy and Astrophysics, University of California, Santa Cruz, CA 95064, USA}

\author{Y.-C.~Pan}
\affiliation{Department of Astronomy and Astrophysics, University of California, Santa Cruz, CA 95064, USA}

\author{A.~L.~Piro}
\affiliation{The Observatories of the Carnegie Institution for Science, 813 Santa Barbara Street, Pasadena, CA 91101}

\author{J.~X.~Prochaska}
\affiliation{Department of Astronomy and Astrophysics, University of California, Santa Cruz, CA 95064, USA}

\author{E.~Ramirez-Ruiz}
\affiliation{Department of Astronomy and Astrophysics, University of California, Santa Cruz, CA 95064, USA}
\affiliation{DARK, Niels Bohr Institute, University of Copenhagen, Blegdamsvej 17, 2100 Copenhagen, Denmark}

\author{A.~Rest}
\affiliation{Space Telescope Science Institute, 3700 San Martin Drive, Baltimore, MD 21218}
\affiliation{Department of Physics and Astronomy, The Johns Hopkins University, 3400 North Charles Street, Baltimore, MD 21218, USA}

\author{C.~Contreras}
\affiliation{Las Campanas Observatory, Carnegie Observatories, Casilla 601, La Serena, Chile}

\author{N.~Morrell}
\affiliation{Las Campanas Observatory, Carnegie Observatories, Casilla 601, La Serena, Chile}

\author{C.~Rojas-Bravo}
\affiliation{Department of Astronomy and Astrophysics, University of California, Santa Cruz, CA 95064, USA}

\author{J.~D.~Simon}
\affiliation{The Observatories of the Carnegie Institution for Science, 813 Santa Barbara Street, Pasadena, CA 91101}

\begin{abstract}
  We discovered Swope Supernova Survey 2017a (SSS17a) in the
  LIGO/Virgo Collaboration (LVC) localization volume of GW170817, the
  first detected binary neutron star (BNS) merger, only 10.9 hours
  after the trigger.  No object was present at the location of SSS17a
  only a few days earlier, providing a qualitative spatial and
  temporal association with GW170817.  Here we quantify this
  association, finding that SSS17a is almost certainly the counterpart
  of GW170817, with the chance of a coincidence being $\le${}$9 \times
  10^{-6}$ (90\% confidence).  We arrive at this conclusion by
  comparing the optical properties of SSS17a to other known
  astrophysical transients, finding that SSS17a fades and cools faster
  than any other observed transient.  For instance, SSS17a fades
  $>$5~mag in $g$ within 7~days of our first data point while all
  other known transients of similar luminosity fade by $<$1~mag during
  the same time period.  Its spectra are also unique, being mostly
  featureless, even as it cools.  The rarity of ``SSS17a-like''
  transients combined with the relatively small LVC localization
  volume and recent non-detection imply the extremely unlikely chance
  coincidence.  We find that the volumetric rate of SSS17a-like
  transients is $\le${}$1.6 \times 10^{4}$~Gpc$^{-3}$~year$^{-1}$ and
  the Milky Way rate is $\le$0.19 per century.  A transient survey
  designed to discover similar events should be high cadence and
  observe in red filters.  The LVC will likely detect substantially
  more BNS mergers than current optical surveys will independently
  discover SSS17a-like transients, however a 1-day cadence survey with
  LSST could discover an order of magnitude more events.
\end{abstract}



\defcitealias{Drout14}{D14}

\section{Introduction}\label{s:intro}
The transient sky is filled with dozens of different observationally
distinct classes of explosive transients.  Although most fall into
three broad classes, Type Ia, Type II, and Type Ib/c supernovae
(SNe~Ia, II, and Ib/c, respectively), there are many additional
classes of ``exotic'' or ``peculiar'' transients including Ca-rich
SNe, luminous red novae, luminous SNe~IIn, pair-instability SNe, SN
impostors, SNe Iax, SN~2006bt-like SNe, and tidal disruption flares
\citep[e.g.,][]{Smith07:06gy, Berger09:ngc, Gal-Yam09, Foley10:06bt,
  Foley13:iax, Perets10:05e, Gezari12, Quimby11}.  These classes have
diverse observational properties with peak luminosities spanning $>$3
orders of magnitude, timescales ranging from about two weeks to $>$1
year, and spectra that bely significantly different compositions,
ejecta velocities, and ionization states.  Quite simply, the transient
sky is extremely diverse.

Despite this diversity, known transients still do not fill the entire
parameter space of peak luminosity--timescale--velocity space
\citep[e.g.,][]{Villar17}.  Instead, there are clear correlations
between these parameters that can be explained by most transients
coming from a limited set of progenitor stars (generally white dwarfs
or massive stars), energy sources (generally the decay of $^{56}$Ni, H
recombination, or interaction between a shock and circumstellar
material), and emission mechanisms (thermal or synchrotron).
Therefore new progenitors and/or energy sources are likely necessary
to produce a transient with truly unique observational properties.

At 12:41:04 on 17 August 2017 (all times are UT), the advanced LIGO
detector \citep{Abbott:AdvLIGO} at Hanford measured a
gravitational-wave transient (later labeled GW170817
\citep{Abbott17:detection}); the same event was then detected in the
advanced LIGO detector data at Livingston and identified by the LIGO
Scientific Collaboration and the Virgo collaboration (LVC) as a binary
neutron star (BNS) merger \citep{GCN21505}.  Nearly simultaneously,
Fermi and INTEGRAL detected a short GRB (sGRB), GRB170817A, that was
spatially coincident with GW170817 \citep{GCN21507}. Later, the full
preliminary three-instrument skymap using also the data of the
Advanced Virgo detector \citep{Acernese17:AdvVirgo} was given
\citep{GCN21527} at 23:54:40 UTC.  Only 10.9 hours after the LVC
trigger, we discovered an optical transient, Swope Supernova Survey
2017a (SSS17a; also known as DLT17ck \citealt{GCN21531} and
AT~2017gfo), spatially coincident with GW170817 \citep{GCN21529,
  Coulter17}.  For the following three weeks, we monitored SSS17a in
the optical and near-infrared \citep{Coulter17, Drout17, Shappee17}.

It has been assumed that SSS17a is the electromagnetic (EM)
counterpart to GW170817.  Here we determine the uniqueness of SSS17a
compared to other known transients.  In Section~\ref{s:comp}, we
compare the optical light curves and spectra of SSS17a to other known
transients.  In Section~\ref{s:like}, we determine the likelihood of
SSS17a being associated with GW170817.  We discuss SSS17a in the
context of current and future surveys in Section~\ref{s:disc} and
conclude in Section~\ref{s:conc}.

In this Letter, we assume the Tully-Fisher distance to NGC~4993, the
host galaxy of SSS17a, of $D = 39.5$~Mpc \citep{Freedman01} and no
host-galaxy reddening for SSS17a \citep[consistent with non-detection
of Na~D absorption;][]{Shappee17}.  We assume $H_{0} =
73.2$~km~s$^{-1}$~Mpc$^{-1}$ \citep{Riess16:h0}, $\Omega_{m} = 0.3$,
and $\Omega_{\Lambda} = 0.7$.


\section{Photometric and Spectroscopic Comparisons}\label{s:comp}

Immediately after the discovery of SSS17a \citep{Coulter17}, we
initiated a follow-up campaign to obtain near-nightly 14-band
ultraviolet/optical/NIR photometry \citep{Coulter17, Drout17} and
optical spectra \citep{Shappee17}.  \citet{Kilpatrick17} analyze these
data in the context of kilonova models.  Here we compare to other
known transients with no model assumptions.

\begin{figure*}
\begin{center}
\includegraphics[angle=0,width=3.2in]{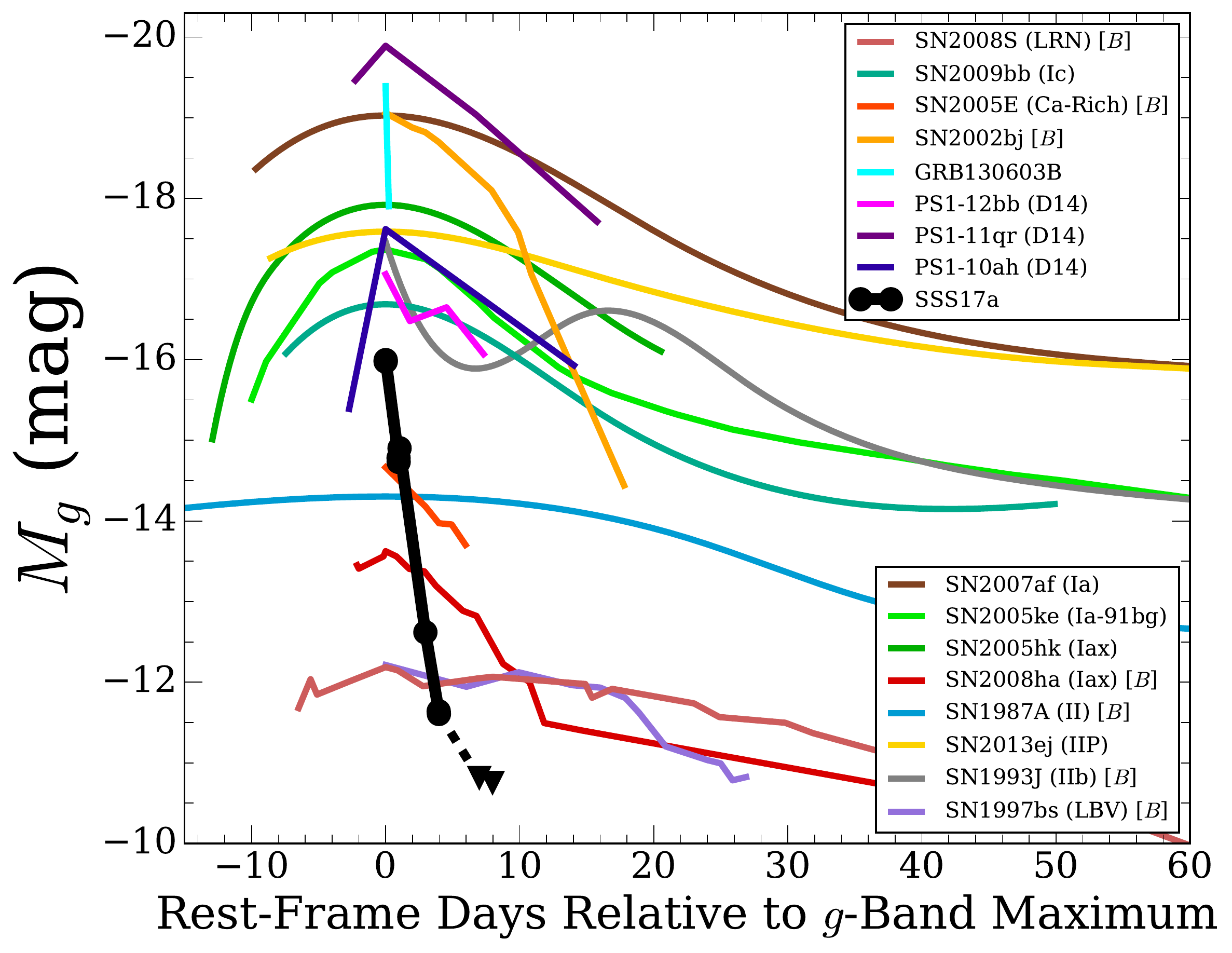}
\includegraphics[angle=0,width=3.2in]{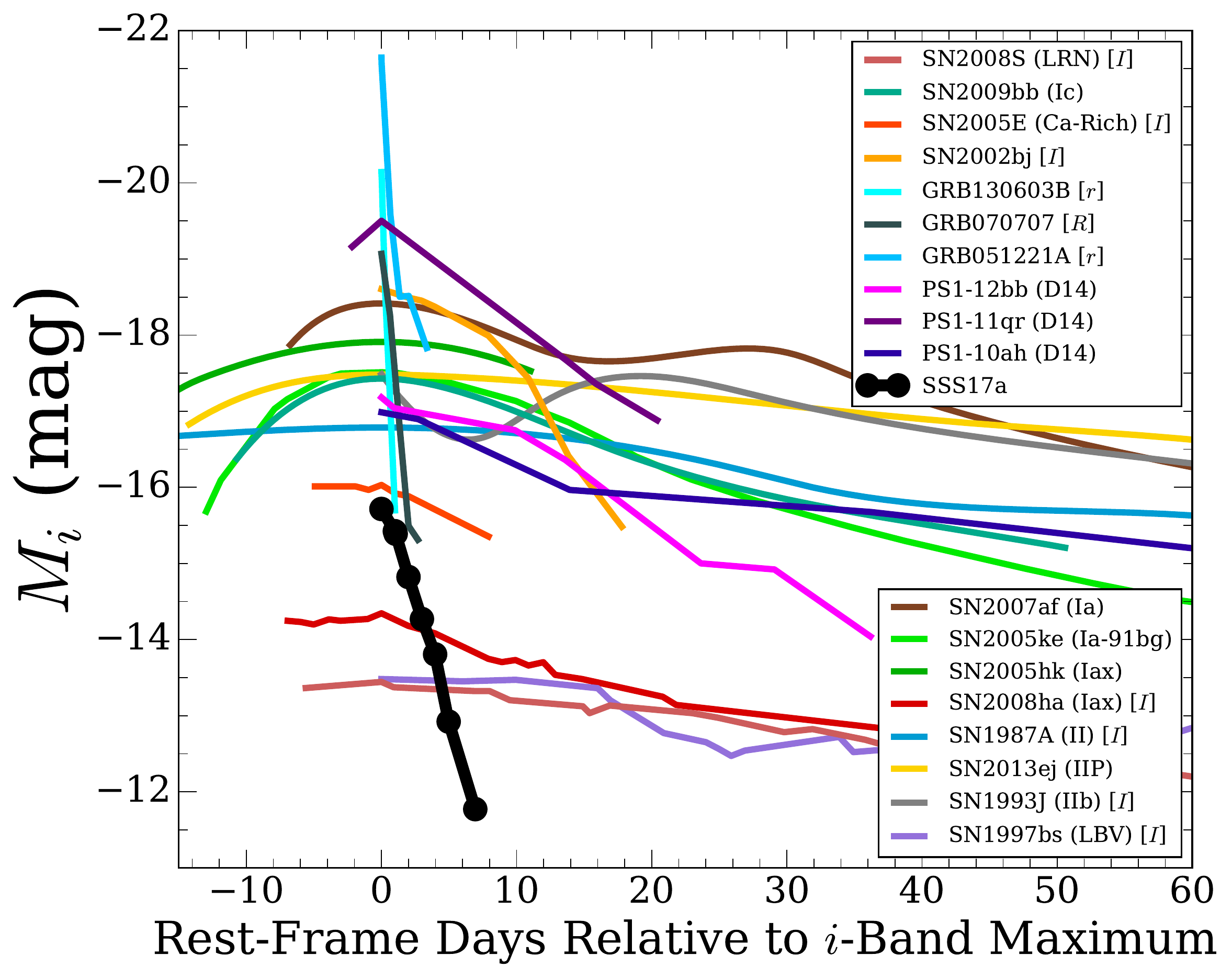}
\caption{Absolute magnitude light curves of SSS17a (black curves) in
  $g$ (left panel) and $i$ (right panel) bands.  Also shown for
  comparison are representative light curves for SNe~Ia (both
  typical-luminosity and a representative of the SN~1991bg-like
  class), Ic, IIP, and IIb, as well as peculiar transients,
  SN~II~1987A, SN~Iax~2005hk, SN~Iax~2008ha, Ca-rich SN~2005E,
  fast-evolving SN~Ib~2002bj \citep{Poznanski:02bj}, luminous red nova
  SN~2008S \citep{Prieto08, Botticella09}, multiple members of the
  \citetalias{Drout14} sample of fast-evolving, blue transients, and
  several sGRBs. All light curves are presented in the observer frame.
  SSS17a evolves significantly faster than any non-GRB comparison
  object.  SSS17a is slower than sGRB afterglows and significantly
  less luminous at peak. We use the $B$ and $I$-band light curves for
  all comparisons except SN 2013ej, SN 2005ke, GRB and D14 objects
  (where we use $g/i$). Luminosities for SN 2002bj, SN 2005E, GRB and
  D14 objects assume Hubble distances while the rest are independently
  measured.}\label{f:abs}
\end{center}
\end{figure*}

Figure~\ref{f:abs} presents our $g$- and $i$-band absolute magnitude
($M_{g}$ and $M_{i}$, respectively) light curves.  Also displayed are
the absolute magnitude light curves (in similar bands) of
representative SNe~Ia, low-luminosity SN~1991bg-like SNe~Ia, and
SNe~Ic, IIP, and IIb.  SSS17a is clearly very different from these
classes, being less luminous at peak ($M_{g, \rm{~peak}} = -16.0$~mag;
$M_{i, \rm{~peak}} = -15.7$~mag) and fading much faster than all
typical classes.  We also display SN~1987A, which is low luminosity
but very slow, the ``typical'' SN~Iax~2005hk \citep{Phillips07}, the
low-luminosity SN~Iax~2008ha \citep{Foley09:08ha, Foley10:08ha}, the
Ca-rich SN~2005E \citep{Perets10:05e}, three fast-evolving, blue
\citet[hereafter \citetalias{Drout14}]{Drout14} transients, and
several sGRB afterglows.  SSS17a has a peak luminosity somewhat
similar to SN~1987A, Ca-rich SNe, and SNe~Iax, and is lower luminosity
than the \citetalias{Drout14} transients and significantly fainter
than sGRBs.  While SNe~2005E, 2008ha, and the \citetalias{Drout14}
transients all fade quickly relative to common SNe, none fades nearly
as rapidly as SSS17a. On the other hand, all three short-GRB
afterglows, GRB130603B, GRB070707 and GRB051221A \citep{Berger13,
  Piranomonte08, Soderberg06}, decline extremely rapidly during the
time they are visible and are significantly faster than SSS17a.

Although SSS17a is associated with GRB170817A, it is inconsistent with
being an afterglow \citep{Drout17}.  Its broadband spectral-energy
distribution (SED), and its evolution, are consistent with a rapidly
cooling thermal spectrum \citep{Drout17} and its detailed spectral
evolution is inconsistent with a synchrotron spectrum
\citep{Shappee17}.  As no other sGRB is known to have a non-thermal
optical SED, SSS17a is clearly not simply a GRB afterglow.
Furthermore, significantly more energy is radiated from SSS17a than
originally from GRB170817A \citep{Murguia-Berthier17}, even making the
phrase ``afterglow'' a misnomer.  Therefore, we focus our subsequent
comparisons on other transients that are discovered by optical
surveys.

\begin{figure*}
\begin{center}
\includegraphics[angle=0,width=3.2in]{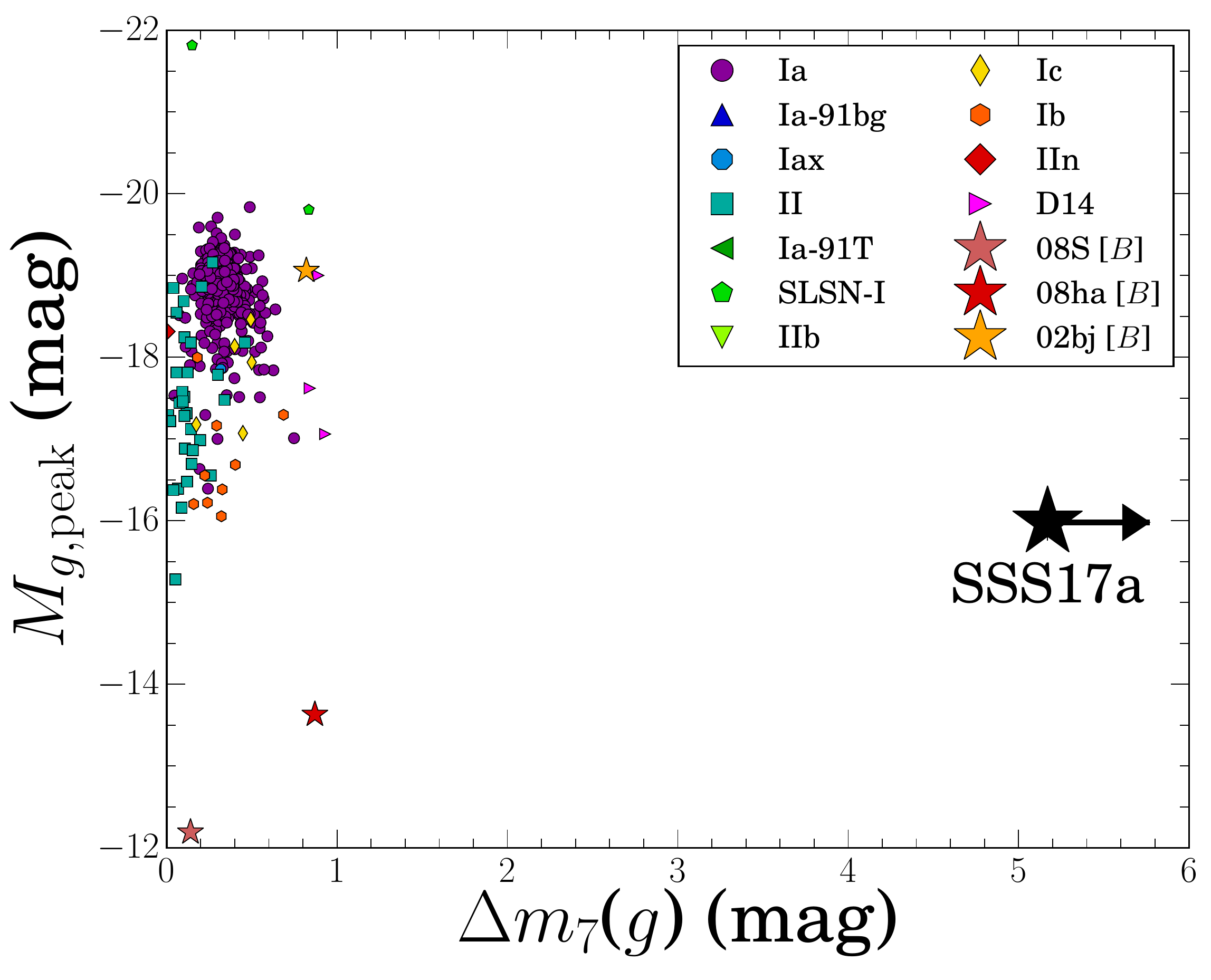}
\includegraphics[angle=0,width=3.2in]{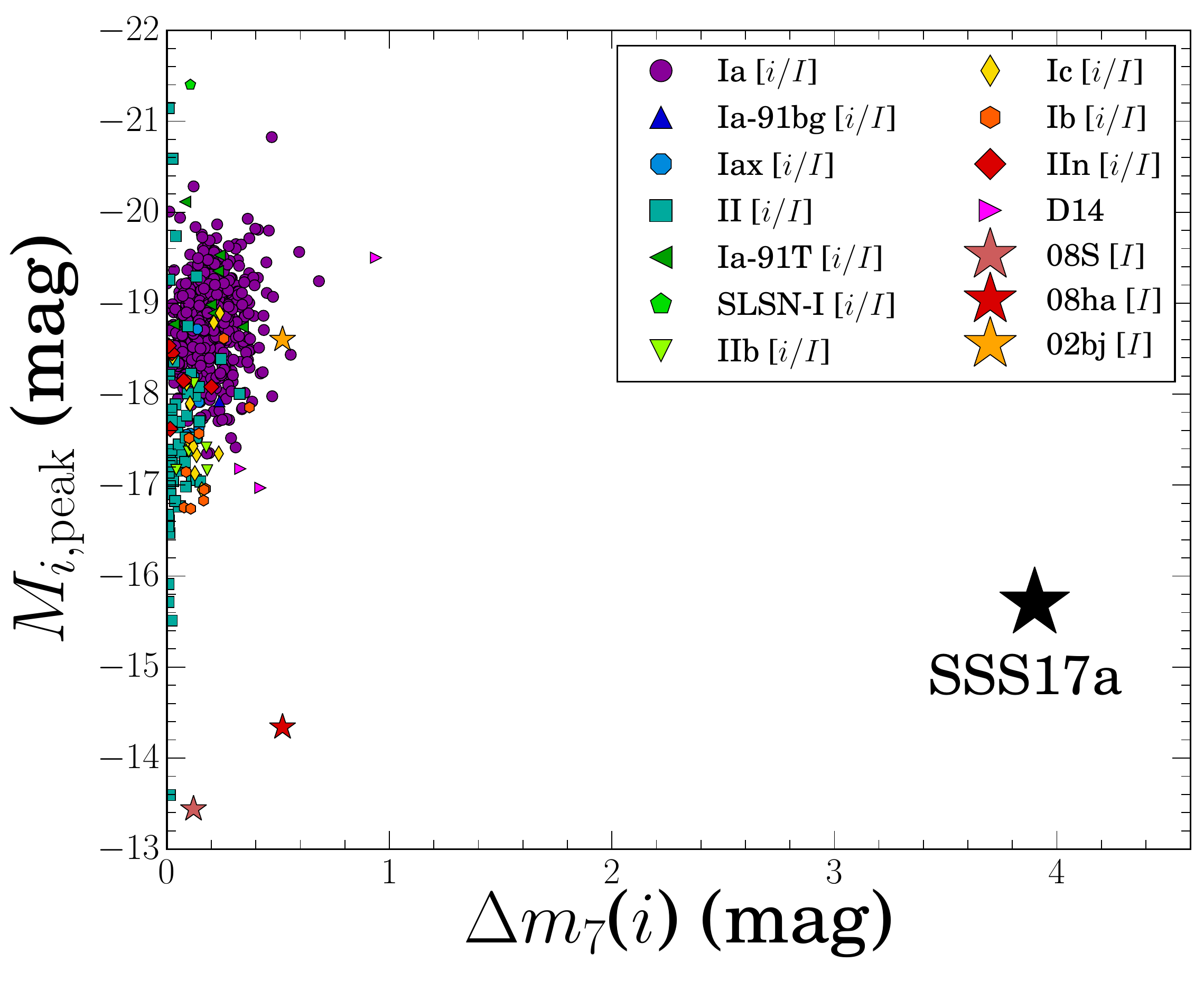}
\caption{Comparison between peak luminosity and the magnitude decline
  after 7~days from peak ($\Delta m_{7}$) in the $g$ (left panel) and
  $i/I$ (right panel) bands for SSS17a (black star) and other
  transient objects.  Transients of a particular class are shown with
  a similar color.  We correct for Milky Way extinction, but not for
  any host-galaxy extinction.  While SSS17a declines by $>$5.17~mag in
  $g$ within the first 7~days after peak, no other transient declines
  by more than 1~mag. Luminosities assume either Hubble distances for
  $z > 0.015$ or independently measured distances for closer
  objects.}\label{f:dm7}
\end{center}
\end{figure*}

The decline of transients are often described by the magnitude decline
from peak over 15~days, $\Delta m_{15}$ \citep{Phillips93}.  However,
SSS17a faded so quickly in the optical that we have no measurements
near 15~days after peak.  Instead, we use $\Delta m_{7}$, finding
$\Delta m_{7} (g) > 5.2$ and $\Delta m_{7} (i) = 3.90 \pm 0.18$~mag
(with the assumption that our first data point corresponds to peak in
each band).  Figure~\ref{f:dm7} displays the peak luminosity and
$\Delta m_{7}$ for SSS17a and all other transients with sufficient
data in the Open Supernova Catalog \citep{Guillochon16}.  We note that
we do not correct the peak absolute magnitude of these objects for
host-galaxy extinction.  Although this is a biased sample that depends
on which objects were followed, it is still informative and peculiar
transients are over-represented relative to a magnitude-limited
sample. The median decline rates for all comparison transients are
$\Delta m_{7} (g) = 0.31$~mag and $\Delta m_{7} (i) = 0.18$~mag,
respectively.  The second-fastest transient in $i$-band, after SSS17a,
is PS1-11qr, a \citetalias{Drout14} transient with $\Delta m_{7} (i) =
0.94$~mag.  Figure~\ref{f:dm7} is a striking representation of how
unique SSS17a is relative to previously observed transients, including
those previously described as ``exotic.''

\begin{figure*}
\begin{center}
\includegraphics[angle=0,width=2.3in]{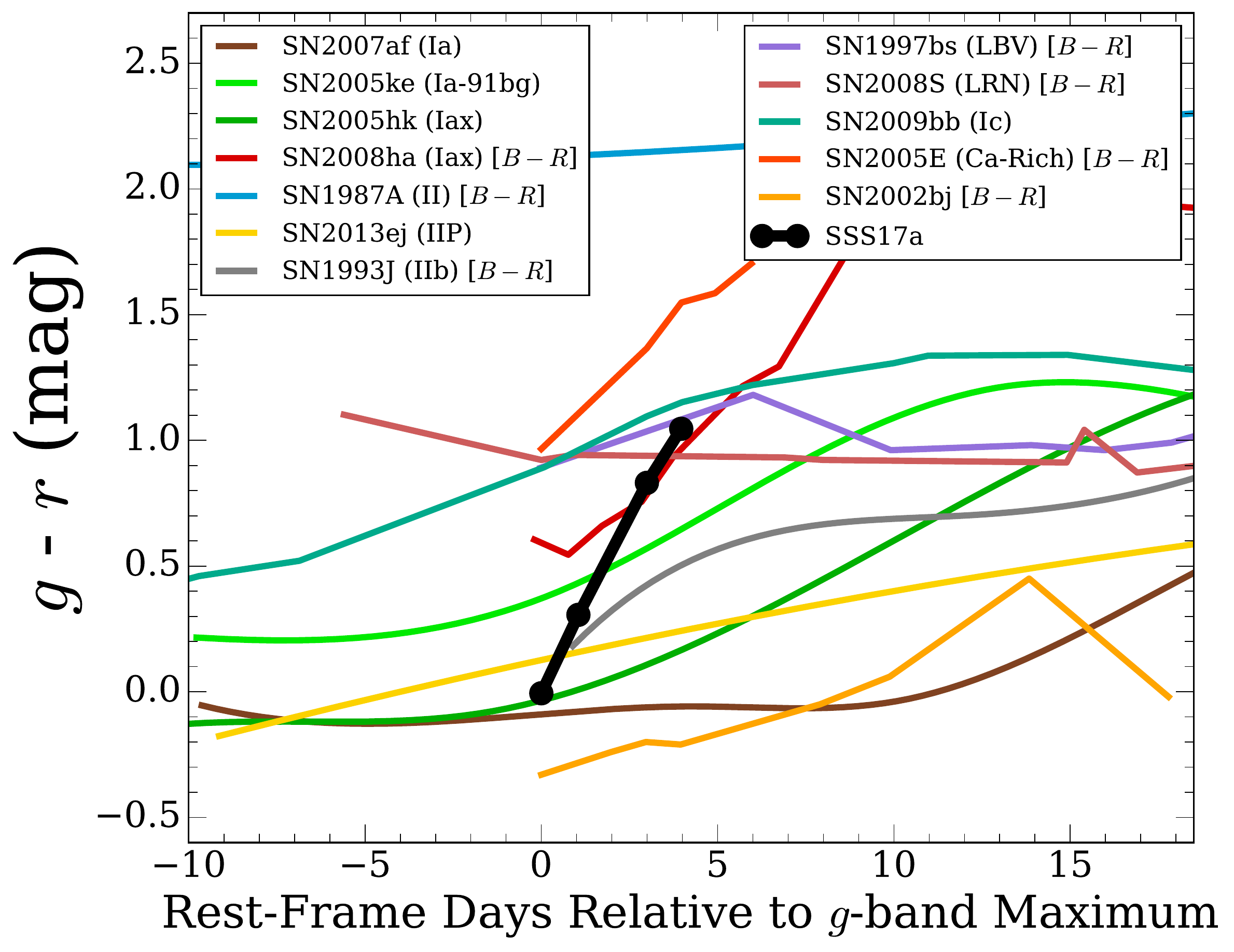}
\includegraphics[angle=0,width=2.3in]{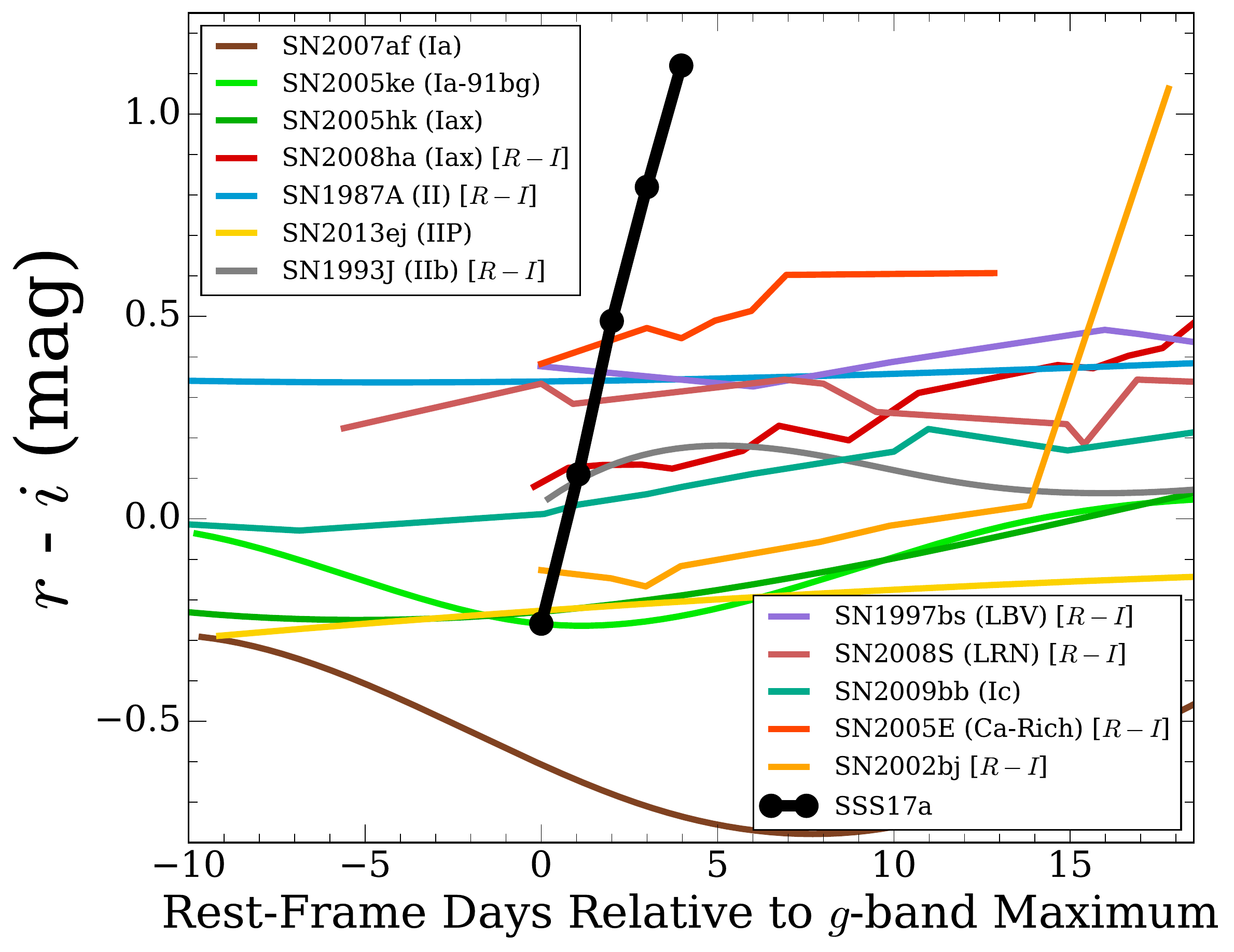}
\includegraphics[angle=0,width=2.3in]{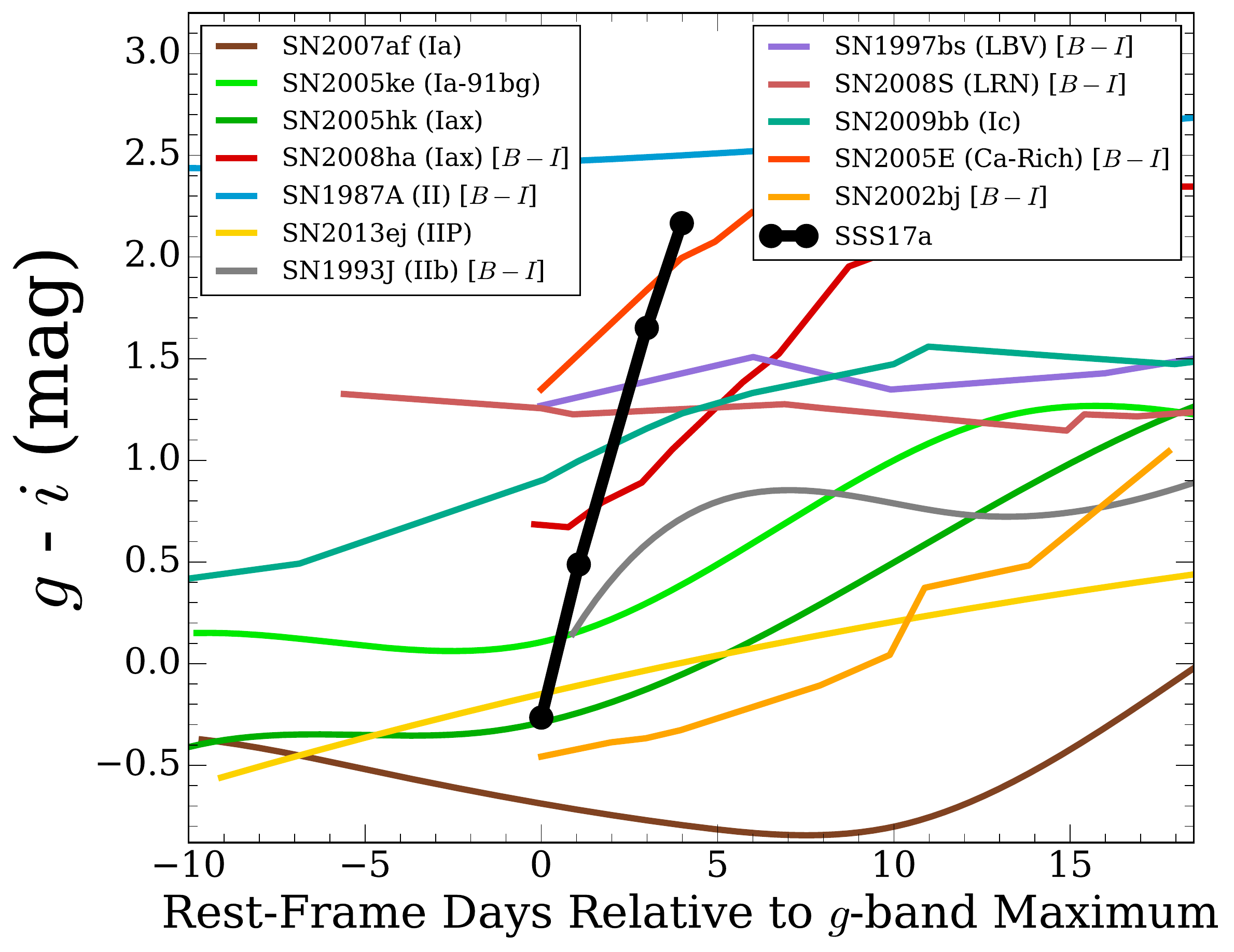}
\caption{Color curves of SSS17a (black curves) in $g-r$ (left panel),
  $r-i$ (middle panel), and $g-i$ (right panel).  Also shown are
  comparison objects from other transient classes.  See
  Figure~\ref{f:abs} for details.  We correct for Milky Way
  extinction, but not for any host-galaxy extinction.  While SSS17a
  has a similar $g-r$ color evolution as many objects, its dramatic
  $r-i$ color evolution is unprecedented, where it starts with a color
  near 0.1~mag at peak and quickly evolves to $r-i \approx 1.2$~mag
  only 4~days later.}\label{f:col}
\end{center}
\end{figure*}

In addition to the fast fading of SSS17a, its color also evolves
extremely quickly.  Figure~\ref{f:col} displays $g-r$, $r-i$, and
$g-i$ color curves of SSS17a and comparison objects.  At peak SSS17a
has colors very similar to those of other transients.  However, only a
few days later, SSS17a has significantly redder colors, with its $r-i$
color changing from $-0.26$~mag at peak to 1.12~mag 4.0~days later.
This dramatic color evolution is consistent with the overall
temperature evolution showing that SSS17a cooled from $\sim$10,000~K
at peak to 5100~K a day later, to about 2500~K about a week after peak
\citep{Drout17}.  Such color/temperature evolution has never been seen
in any other transient.

\begin{figure}
\begin{center}
\includegraphics[angle=0,width=3.2in]{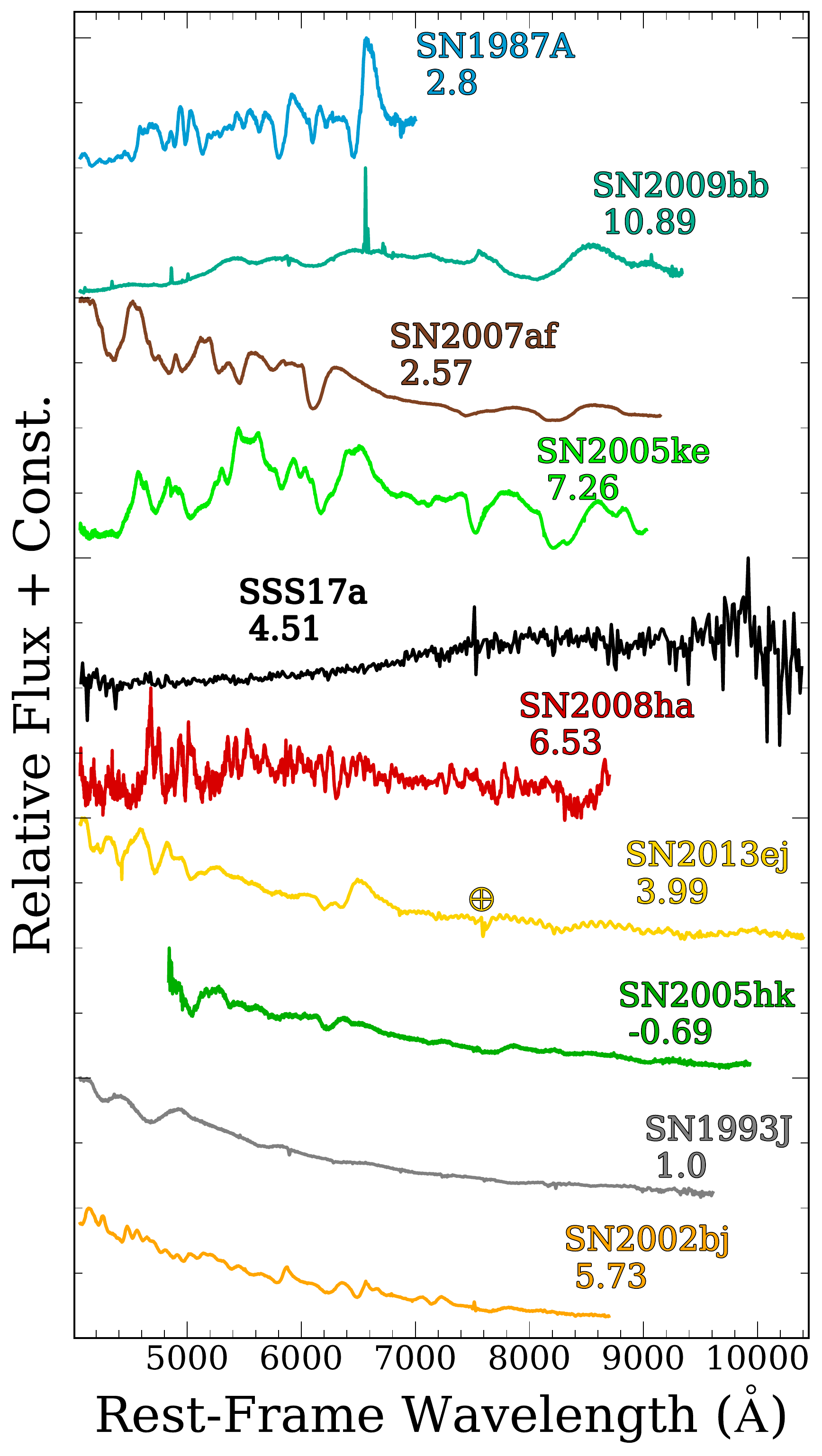}
\caption{Optical spectrum of SSS17a (black curve) 4.5~days after the
  LVC trigger.  Spectra of other transient objects are shown for
  comparison.  Telluric absorption features are marked with an
  `$\earth$' symbol.  No other SN has spectra similar to SSS17a at
  +4.5~days, and while there are some objects with spectra similar to
  SSS17a at +11~hours, none evolves nearly as quickly as
  SSS17a.}\label{f:spec}
\end{center}
\end{figure}

\citet{Shappee17} performed a detailed comparison of the SSS17a
spectral sequence to other transients.  The earliest spectra are blue
and featureless, similar to several other young transients.  However,
SSS17a already has distinct spectra 1~day later.  Figure~\ref{f:spec}
displays the +4.5-day spectrum of SSS17a compared to other transients
at similar phases.  SN~2002bj is the fastest evolving SN of the
comparison transients.  However, there are prominent He and C lines
superimposed on a blue continuum \citep{Poznanski:02bj} at 5.73~days
after $B$-band maximum.  The color and luminosity of SSS17a are
somewhat similar to those of SN~1987A, but the evolutionary timescale
of its spectral and photometric properties is orders of magnitude
shorter.  All transient spectra, except for that of SSS17a, exhibit
relatively narrow spectral features.  SSS17a has a very broad
``feature'' from $\sim$8000--10,000~\AA, which may be the result of
combining physically distinct ejecta components \citep{Shappee17}.
Regardless, SSS17a has spectra that are significantly different from
those of other previously known transients.

\section{Likelihood of SSS17a Being Associated with GW170817}\label{s:like}

SSS17a has truly unique observational properties relative to all other
known transients.  Qualitatively, the unprecedented nature of SSS17a
combined with the spatial and temporal coincidence provides strong
evidence that SSS17a is the EM counterpart to GW170817.  Here we
quantify this association and update our initial estimates \citep{GCN21557}.

To perform this calculation, we must first determine the relative rate
of transients similar to SSS17a and other transients in transient
surveys.  We note that in the surveys relevant for this calculation,
no GRB afterglow has been detected (although see \citealt{Cenko15}).
Each survey observes to a limiting magnitude, which provides a volume
in which a particular transient can be discovered.  Since SSS17a fades
so quickly, we must be careful to determine the volume in which a
survey would detect SSS17a on at least two separate epochs.  For these
calculations, we follow the procedure outlined by \citet{Foley13:iax}.

One of the most prolific transient surveys is the Lick Observatory
Supernova Search \citep[LOSS;][]{Filippenko01}, which has performed a
galaxy-targeted SN search for two decades.  \citet{Leaman11} presented
a sample of 726 LOSS-discovered SNe.  Although their cadence varied
over the entire survey, the closest galaxies were generally observed
at a cadence of $\sim$5~days.  LOSS has an unfiltered search, but its
CCD response is similar to $R$ band \citep{Filippenko:loss}.  With a
typical limiting magnitude of 18.5, LOSS would detect SSS17a only to a
distance modulus of $\mu_{\rm max} = 30.9$~mag (roughly 15~Mpc).
There are only 12 SNe in the \citet{Leaman11} sample within this
volume.  However, since 2008, which marks the end of the
\citet{Leaman11} sample, an additional 46 SNe were detected within
this volume.  While many of these objects may have been detected by
low-cadence ($>$7 days) surveys, new surveys tend to have higher
cadence ($<$3 days) than LOSS ($\sim$5~days).  We therefore use a
total sample of 58 SNe as occurring within the SSS17a detection volume
as part of surveys that could have detected a transient similar to
SSS17a.

Since we do not have detailed logs of each survey, it is difficult to
precisely determine the correction factor necessary to account for the
difficulty of observing SSS17a.  Such a factor is necessary to account
for the fact that while SSS17a could be detected within the volume
determined above, doing so requires no significant reddening, no long
periods of bad weather, and the transient occurring when it is not
behind the Sun (other nearby SNe can be discovered in this volume over
many months).  Sun constraints and weather alone reduce the efficiency
of discovering the most-distant SSS17a-like objects by a factor of
$\sim$2.  To account for unknown inefficiencies in different surveys,
we choose a conservative correction factor of 4, which means that for
every 4 SSS17a-like objects that occur in the volume, we would only
detect 1.  As a comparison, \citet{Foley13:iax} used a correction
factor of 2 for the closest SNe~Iax.

Using Poisson statistics, we determine the rate of SSS17a-like
transients necessary such that 90\% of the time, we would have
expected to detect at least one SSS17a-like transient in previous
surveys, finding that the SSS17a-like transient rate is at most 0.16
that of all other SNe.  That is, if the fraction of SSS17a-like
objects were higher than $f_{\rm SSS17a} = 0.16$ (16\%) that of
overall SN population with at least its luminosity, we would have
expected LOSS or other surveys since 2008 to be more likely to detect
at least one SSS17a-like object with 90\% confidence.

The LVC localizes the source at a luminosity distance of $40 \pm 7$
Mpc within a sky region of 31 deg$^{2}$ (at 90\% confidence)
\citep{Abbott17:detection}, corresponding to a volume of
104~Mpc$^{3}$.  Using the \citet{Li11:rate2} total volumetric SN rate,
we expect a SN rate in the LVC localization region of $R_{\rm LVC} =
0.010$~SNe~year$^{-1}$.

Finally, we note that there was no detection of a transient at the
location of SSS17a $t_{nd} = 2.0$, 21.0, and 111.8~days before its
detection by ASAS-SN, DLT40, and {\it Hubble Space Telescope}
\citep{Shappee14, Shappee17:asassn, GCN21533, GCN21579, GCN21536, Pan17}.
Although it is unclear how bright SSS17a could have been two days
before detection, it would likely be detectable in the ASAS-SN image
if we simply extrapolate back from the time of detection.  Ignoring
the ASAS-SN data, SSS17a was at most 21.0~days old at the time of
discovery.  These non-detections indicate that SSS17a was relatively
young when discovered.

We can combine the above individual measurements to determine the
frequency of SSS17a-like objects in a volume consistent with the LVC
localization and within a specific time frame,
\begin{equation}
  P_{\rm chance} \le f_{\rm SSS17a} \times R_{\rm LVC} \times t_{\rm nd},
\end{equation}
where $P_{\rm chance}$ is the probability that SSS17a is a spatial and
temporal chance coincidence with GW170817.  Using the values
calculated above, we find $P_{\rm chance} \le 9 \times 10^{-6}$ at
90\% confidence (or $1 \times 10^{-4}$ using the DLT40
non-detection) .  Therefore, it is extremely unlikely that SSS17a is
unassociated with GW170817 and SSS17a is almost certainly the
optical/NIR counterpart to GW170817.  Adding further confidence that
SSS17a is the counterpart to GW170817, \citet{Coulter17} detected no
other transients in its search covering 2 weeks after the trigger and
covering 95.3\% of the probability region (when combined with galaxy
properties).


\section{Discussion }\label{s:disc}

SSS17a is a truly unique object.  Since similar objects have not yet
been discovered through searches unguided by a GW alert, it is
difficult to determine the rate of such objects.  Nevertheless, we
have placed a limit on the rate of SSS17a-like objects relative to the
SN rate in a volume-limited survey, $f_{\rm SSS17a} \le 0.16$.  Using
the \citet{Li11:rate3} volumetric SN rate, we determine the rate of
SSS17a-like events to be $\le${}$1.6 \times
10^{5}$~Gpc$^{-3}$~year$^{-1}$ and a rate in the Milky Way of
$\le$0.46 SSS17a-like transients per century.

Current population synthesis models predict the BNS merger rate in the
Milky Way to be 0.0024 per century \citep{Chruslinska17}, two orders
of magnitude below our estimate. The upper limit on the volumetric BNS
merger rate from the LIGO O1 observing run is $12.6 \times
10^{3}$~Gpc$^{-3}$~year$^{-1}$ \citep{LIGO:ul}, comparable to our
current estimate using independent data.

However, all values above required a GW alert.  If we want to
determine an independent measurement of the rate of BNS mergers and
kilonovae from EM observations alone, we must have a survey that
independently discovers such objects.  Based on the observational
properties of SSS17a, we can suggest a survey design to maximize their
detection and thus the rate determination.

SSS17a fades quickly, so a high-cadence ($<$3-day) survey is key.
Since SSS17a faded by $>$1~mag in $g$ between 0.5 and 1.5~days after
the LVC trigger, but only 0.3~mag during the same time in $i$, with
$g-i \approx 0.6$~mag at $+1.5$~days, a survey in a redder band is
clearly preferred.  In fact, SSS17a ``only'' faded by 2.3~mag between
$+1.5$ and $+8.5$~days in $z$ band, indicating that a $z$-band search
has distinct advantages over bluer bands.

In principle, one can now use the kilonova rate to determine the BNS
merger rate.  If one wants to discover a comparable number of
kilonovae with an optical survey as LVC will discover BNS mergers, the
optical survey must cover the same volume--time as LVC.  In 2018-2019
advanced LIGO and Advanced Virgo are expected reach BNS ranges of
120-170 Mpc and 65-115 Mpc \citep{Abbott:Local} respectively,
corresponding to a maximum volume of 0.021~Gpc$^{3}$.  Assuming a 50\%
duty cycle, LVC should survey 0.010~Gpc$^{3}$-year every year of
operation.  An optical survey might miss only $\sim$10\% of all
possible kilonovae to weather/instrument failures if it has an
adequately deep, high-cadence survey.  While LVC will monitor the
entire sky (although to varying distances depending on sky position),
an optical search would likely only survey $\sim$35\% of the sky at
any time because of the Sun, declination limits, and the Milky Way.
Given these constraints, the optical survey must reach a distance of
$\sim$200~Mpc, or a distance modulus of $\mu = 36.5$~mag, to have the
same volume--time as LVC.  Thus, a nightly, ``all sky'' $z$-band
($i$-band) survey would need to reach a limiting magnitude of 21.1~mag
(21.3~mag) to be competitive with LVC.  Depending on the specifics of
the telescope, site, and camera, an $i$-band survey may be more
efficient than a $z$-band survey.

Currently, no survey reaches this goal.  The most comparable are ATLAS
\citep{Shanks15}, which generally reaches a $o$-band (roughly $i+z$)
limiting magnitude of 19.5 and covers a large fraction of the sky on a
few-day cadence, and soon the Zwicky Transient Facility
\citep{Bellm14}, which will reach a limiting magnitude of 20.4 in $r$
band (where the limiting magnitude needs to be 21.4 to be competitive
with LVC).

Interestingly, LSST is expected to have a limiting magnitude of 23.3
in $z$ for a 30-second exposure \citep{LSST}.  For such a deep
exposure, one could potentially run a survey to 550~Mpc ($z = 0.11$),
which would probe a volume--time that is a factor of 24 larger than
LVC at design sensitivity.  If SSS17a is typical for all BNS mergers,
LSST should be able to produce a significantly more constraining BNS
merger rate than LVC until GW detectors have a BNS horizon of
$\sim$470~Mpc.


\section{Conclusions}\label{s:conc}

We have compared the optical properties of SSS17a, the first EM
counterpart to a gravitational wave source, to other known transients.
While its luminosity is similar to some SNe and other transients, all
other properties of SSS17a are distinct from known classes of
transients.  In particular, SSS17a fades significantly faster than all
non-relativistic transients, has a dramatic blue to red color
evolution in only a week, and has relatively featureless spectra, even
after it has cooled significantly.  We conclude that SSS17a is unique
among all known transients.

Based on the uniqueness of SSS17a and its spatial/temporal coincidence
with GW170817, we determine that there is $\le${}$9 \times 10^{-6}$
chance that SSS17a and GW170817 are physically unrelated at 90\%
confidence.  Therefore, SSS17a is almost certainly the EM counterpart
to GW170817.

As no transient discovered by an optical survey is similar to SSS17a,
we can limit the relative rate of SSS17a-like transients, finding that
they have a rate at most 16\% that of all other SNe combined.  We also
limit the volumetric and Milky Way rate of SSS17a-like events to be
$\le${}$1.6 \times 10^{4}$~Gpc$^{-3}$~year$^{-1}$ and $\le$0.19 per
century, respectively.  While our limits are not competitive with
population synthesis expectations, they are comparable to previous
constraints from LIGO.

Because SSS17a-like transients are relatively faint, fast fading, and
relatively rare, it is unlikely that any current surveys will be able
to detect more such events than LVC will be able to detect BNS
mergers.  However, LSST, if it performed a 1-day cadence search, could
detect an order of magnitude more SSS17a-like events than GW detectors
would detect BNS mergers.  Such a survey would produce the tightest
constraints on the BNS merger rate unless GW detectors significantly
improved their sensitivity.

Now that we have measured the optical properties of the EM
counterparts of a BNS merger, we can appropriately design observing
programs to both follow-up GW detections and to discover such events
without a GW trigger.  Future discoveries will unveil the diversity of
EM counterparts to BNS mergers.  With a sample of such events, we will
be able to determine if SSS17a is also unique among its peers.

\begin{acknowledgments} 

\bigskip

We thank the editor, F.\ Rasio, for expediting the review process.  We
thank the anonymous referee for an extremely quick and constructive
report.

We thank the University of Copenhagen, DARK Cosmology Centre, and the
Niels Bohr International Academy for hosting D.A.C., R.J.F., A.M.B.,
E.R., and M.R.S.\ during the discovery of GW170817/SSS17a.  R.J.F.,
A.M.B., and E.R.\ were participating in the Kavli Summer Program in
Astrophysics, ``Astrophysics with gravitational wave detections.''
This program was supported by the the Kavli Foundation, Danish
National Research Foundation, the Niels Bohr International Academy,
and the DARK Cosmology Centre.

The UCSC group is supported in part by NSF grant AST--1518052, the
Gordon \& Betty Moore Foundation, the Heising-Simons Foundation,
generous donations from many individuals through a UCSC Giving Day
grant, and from fellowships from the Alfred P.\ Sloan Foundation
(R.J.F), the David and Lucile Packard Foundation (R.J.F.\ and E.R.)
and the Niels Bohr Professorship from the DNRF (E.R.).
A.M.B.\ acknowledges support from a UCMEXUS-CONACYT Doctoral
Fellowship.
M.R.D. and B.J.S. were partially supported by NASA through Hubble
Fellowship grant HST--HF--51348.001 and HST--HF--51373.001 awarded by
the Space Telescope Science Institute, which is operated by the
Association of Universities for Research in Astronomy, Inc., for NASA,
under contract NAS5--26555.

This paper includes data gathered with the 6.5 meter Magellan
Telescopes located at Las Campanas Observatory, Chile.
This research has made use of the NASA/IPAC Extragalactic Database
(NED) which is operated by the Jet Propulsion Laboratory, California
Institute of Technology, under contract with the National Aeronautics
and Space Administration.

\end{acknowledgments}

\bibliographystyle{aasjournal}
\bibliography{gw}

\end{document}